\documentclass[fleqn,10pt]{wlscirep}
\usepackage[utf8]{inputenc}
\usepackage[T1]{fontenc}
\usepackage{appendix}

\usepackage{color}
\usepackage{amsmath}
\usepackage{amssymb}
\usepackage{graphicx}
\usepackage{babel}
\usepackage[normalem]{ulem}

\title{A quantum walk simulation of extra dimensions with warped geometry}
\author[1,*]{Andreu Anglés-Castillo}
\affil[1]{Universitat de València-CSIC, Departament de Física Teòrica \& IFIC,   Burjassot (València), 46100,  Spain}
\author[1]{Armando Pérez}

\affil[*]{andreu.angles@ific.uv.es}

\begin{abstract}
We investigate the properties of a quantum walk which can simulate
the behavior of a spin $1/2$ particle in a model with an ordinary
spatial dimension, and one extra dimension with warped geometry between
two branes. Such a setup constitutes a $1+1$ dimensional version
of the Randall-Sundrum model, which plays an important role in high
energy physics. In the continuum spacetime limit, the quantum walk
reproduces the Dirac equation corresponding to the model, which allows
to anticipate some of the properties that can be reproduced by the
quantum walk. In particular, we observe that the probability distribution
becomes, at large time steps, concentrated near the "low
energy" brane, and can be approximated as the lowest
eigenstate of the continuum Hamiltonian that is compatible with the
symmetries of the model. In this way, we obtain a localization effect
whose strength is controlled by a warp coefficient. In other
words, here localization arises from the geometry of the model, at
variance with the usual effect that is originated from random irregularities,
as in Anderson localization. In summary, we establish an interesting
correspondence between a high energy physics model and localization
in quantum walks. 
\end{abstract}

\begin{document}

\flushbottom
\maketitle
%
%
\thispagestyle{empty}

\section*{Introduction}

Quantum walks (QWs) constitute an interesting possibility for simulating
physical phenomena from many fields. The discrete time version describes
the motion of a spin $1/2$ particle on a lattice. For instance, by simply incorporating
suitable position-dependent phases on the unitary operator that implements
the time evolution, one can mimic the effects of an external electromagnetic
field \cite{PhysRevLett.111.160601,Arnault2016,di2014quantum,PhysRevA.92.042324,PhysRevA.93.032333,PhysRevA.93.052301,PhysRevA.98.032333,Cedzich2018}.
In the continuum limit (when both the time step and the lattice spacing
tend to zero), the Dirac equation in presence of such fields is recovered.
In an analogous way, the motion of a Dirac particle in presence of
a gravitational field can be simulated by an appropriate choice of
the operator that drives the evolution, either on a rectangular or
other types of lattices \cite{di2014quantum,DebbaschWaves,Arrighi2019}.
Other scenarios include vacuum or matter neutrino oscillations \cite{Molfetta2016,Mallick2017,Jha2020},
and one can even establish some connections to lattice field theories
\cite{PhysRevA.99.032110}.

There is also a different connection of QWs with quantum field theories,
namely the possibility to explore some models which include extra
dimensions, which are only manifested at very high energies. The possibility
of extra dimensions of space was first suggested by Theodor Kaluza
and Oscar Klein \cite{Kaluza1921,Klein1926} seeking an unified
theory of electromagnetic and gravitational fields into a higher dimensional
field, with one of the dimensions compactified. Experimental data
from particle colliders restrict the compactification radius to such
small scales that it becomes virtually impossible to explore these extra dimensions.
 Different ideas have been proposed to overcome this
difficulty, for example the domain wall model introduced by Rubakov
and Shaposhnikov \cite{Rubakov1983}, in which the particle couples
to an external scalar field. The motion of a spin $1/2$ particle moving inside
such a geometry was analyzed in \cite{PhysRevA.95.042112}. In addition
to recovering the corresponding Dirac equation in the continuum limit,
the QW shows, at finite spacetime spacing, localization of the particle
within the brane due to the coupling to the field. 

Spatial localization is an important phenomenon in physics, which
appears within the context of diffusion processes in lattices. It
can arise from random noise on the lattice sites, giving rise to Anderson
localization \cite{Lattices1956} and causing a metal-insulator
transition, but it can also be the consequence of the action of an
external periodic potential (see e.g. \cite{aubry1980analyticity,PhysRevLett.49.833,PhysRevLett.103.013901}).
Similarly, one obtains localization for the 1-dimensional QW when
spatial disorder is included \cite{Joye2010,PhysRevLett.106.180403,Crespi2013a},
non-linear effects \cite{Navarrete-Benlloch2007}, or by the use
of a spatially periodic coin \cite{PhysRevE.82.031122}. The results
in \cite{PhysRevA.95.042112} show, however, that localization can
also appear as a consequence of the interaction with a\textit{ smooth}
external potential, instead of a random, or even periodic, perturbation.

In this paper, we investigate localization effects that arise within
a different context, which is also inspired on high energy physics,
and was originally proposed to address the \textit{hierarchy problem}
(the observed difference between the Higgs mass, and the Planck scale,
in many orders of magnitude), and is commonly referred to as the Randall-Sundrum
model \cite{RSoriginal}. This model assumes an extra dimension which
extends between two \textit{branes} (with a topology that will be
discussed later). Here we consider a simplified version with one ordinary
spatial dimension and one extra dimension, and define a QW that reproduces
the dynamics of a spin $1/2$ particle in the continuum spacetime limit. 

Unlike the Rubakov and Shaposhnikov model, there is no coupling
to an external scalar field. Instead, this model presents a warped geometry along
the extra dimension. As we will show, this curvature is at the root
of a localization effect of the QW towards the second (low energy)
brane. 
The stationary states of the model in the continuum
limit become concentrated close to the low energy brane for high values
of the warp coefficient, which quantifies the strength of the localization. 
The localization of the QW can be analyzed by quantifying
its overlap with these stationary states.
This allows us to tailor the dynamics of
the QW, showing a different behavior as the value of the warp coefficient
is changed. In this way, we arrive at a QW model with a rich phenomenology,
where some properties are inherited from the continuum field theoretic model. 
There is, in this sense, a mutual multidisciplinary
benefit: one can design a QW which simulates an important high energy
physics model. In exchange, the knowledge of the continuum properties
is useful to understand, and to control, the dynamics of the QW in different regimes.

This paper is organized as follows. We first define the Randall-Sundrum
model in $1+1$ spatial dimensions, along with its main properties.
We pay special attention to the stationary states of the Hamiltonian,
which play a crucial role in understanding the dynamics of the proposed
QW. Next, we define a QW which allows to recover the dynamics
of the Randall-Sundrum model for a spin $1/2$ particle, and we study
its phenomenology. 
Namely, we show that the distribution probability,
as well as the expected value of the position along the extra dimension, approaches
the lower brane at large time, and that this approaching proceeds
more slowly for larger values of the warp coefficient, which turns out
to be the main parameter in controlling the dynamics. 
We also analyze
the entanglement entropy between spatial and internal degrees of freedom,
exhibiting a complex behavior as a function of that parameter,
which can be attributed to the different sharpness of the probability
distribution. We finally conclude by collecting and discussing our main results. 

\section*{The Model}

\subsection*{Orbifold $S^{1}/\mathbb{Z}_{2}$ and Background Geometry}

\begin{figure}
\centering
	\includegraphics[width=.4\linewidth]{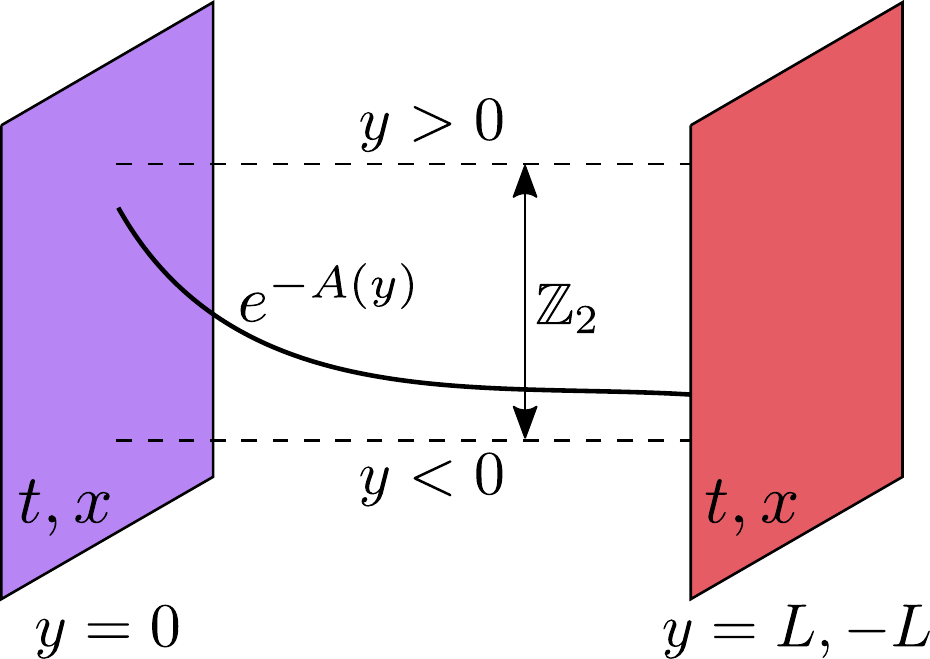}
	\caption{Schematic representation of the extra dimension in the Randall-Sundrum model. }
	\label{fig:Branes} 
\end{figure}

As described in the Introduction, we consider the Randall-Sundrum
model (RSM) \cite{RSoriginal} with a single extra dimension $y$,
together with a 2-dimensional ordinary spacetime, whose coordinates
are denoted by $x^{\mu}=\{t,x\}$. The total spacetime possesses
$D=3$ dimensions. The extra dimension $y$ is compactified on a circle
of radius $R$, and subject to a $\mathbb{Z}_{2}$ symmetry. These
features are captured by the equivalences 
\begin{align}
S^{1}:\;y & \sim y+2\pi R~,\label{eqn:PeriodicCond}\\
\mathbb{Z}_{2}:\;y & \sim-y~,\label{eqn:Z2Cond}
\end{align}
which define the orbifold $S^{1}/\mathbb{Z}_{2}$ describing this
extra dimension. Along the $y$ dimension, the orbifold is a finite
segment with two fixed points at $y=0$ and $y=\pi R\equiv L$. The
RSM assumes that there is a $(D-1)$-brane of ordinary dimensions
at each fixed point, see Fig.~\ref{fig:Branes} for a sketch of the space configuration 
 and the orbifold symmetries. 
The matter fields are supposed to reside on the
brane at $y=L$, which is referred to as the ``visible brane'',
while the brane at $y=0$ is the ``hidden brane''. Both branes contribute
to the bulk background geometry through their tensions, or vacuum energies,
$T_{\mathrm{vis}}$ and $T_{\mathrm{hid}}$ respectively \cite{RSoriginal,BraneTensions}.
The total background action is 
\begin{equation}\label{eqn:ActionBackgroundTotal}
S=\int_{-L}^{L}dy\int dx^{\mu}\sqrt{|g|}\left(2\alpha\mathcal{R}-\Lambda\right)+S_{\mathrm{vis}}+S_{\mathrm{hid}}~,
\end{equation}
where the first term is the usual Einstein-Hilbert action of the total
space, with $\Lambda$ the bulk cosmological constant, $\alpha$ a constant and $|g|$ the absolute value of the metric determinant, while 
\begin{align}
S_{\mathrm{vis}} & =-\int dx^{\mu}\sqrt{|g_{\mathrm{vis}}|}T_{\mathrm{vis}}~,\\
S_{\mathrm{hid}} & =-\int dx^{\mu}\sqrt{|g_{\mathrm{hid}}|}T_{\mathrm{hid}}~,
\end{align}
are the action contributions of the branes tensions, with the induced
metrics $g_{\mathrm{vis}}(x^{\mu})=g(x^{\mu},y=L)$ and $g_{\mathrm{hid}}(x^{\mu})=g(x^{\mu},y=0)$.
To address the hierarchy problem, the following metric was proposed
\begin{equation}
ds^{2}=e^{-2A(y)}\eta_{\mu\nu}dx^{\mu}dx^{\nu}-dy^{2}~,\label{eqn:Metric}
\end{equation}
where $e^{-2A(y)}$ is a \textit{warp factor}, a rapidly changing
function along the additional dimension, and $\eta_{\mu\nu}$ is the
Minkowski metric with signature $(+,-)$. The metric in Eq.~(\ref{eqn:Metric})
obeys Einstein's equations that are obtained from the action (\ref{eqn:ActionBackgroundTotal}): We refer the reader to the Supplementary Material
for the standard computation particularized to
this lower-dimensional spacetime. We also show that, as a consequence
of these equations, the function in the exponent is given by
\begin{equation}
A(y)=k|y|
\end{equation}
where $k$ is the so called \emph{warp coefficient}.

\subsection*{Fermions in the Randall-Sundrum model}

We now focus on the study of spin $1/2$ fermions, whose evolution equation is
the Dirac equation in curved spacetime 
\begin{equation}
(i\gamma^{a}e_{a}^{\mu}D_{\mu}-m)\Psi=0~.\label{eqn:DiracCurved}
\end{equation}
The $\gamma^{a}$ are the Dirac gamma matrices
in a local rest frame, and the covariant derivative is 
\begin{equation}
D_{\mu}=\partial_{\mu}-\frac{i}{4}\omega_{\mu}^{ab}\sigma_{ab}~,\quad\text{with}\quad\sigma_{ab}=\frac{i}{2}[\gamma_{a},\gamma_{b}]~,
\end{equation}
where $\omega_{\mu}^{ab}$ is the spin connection. The \emph{vierbeins} $e_{a}^{\mu}$
allow to express the Dirac matrices in a rest frame, that is, they
perform a change of basis to a non-coordinate system in which the metric
becomes the Minkowski metric 
\begin{equation}
g_{\mu\nu}e_{a}^{\mu}e_{b}^{\nu}=\eta_{ab}~.\label{eqn:VierbeinsDef}
\end{equation}
Equation~(\ref{eqn:DiracCurved}) defines the vector current 
\begin{equation}
j^{\mu}=\sqrt{|g|}e_{a}^{\mu}\overline{\Psi}\gamma_{a}\Psi~,
\end{equation}
whose conservation $\partial_{\mu}j^{\mu}=0$ imposes the normalization
condition 
\begin{equation}
\int dx^{\mu}\sqrt{|g|}e_{0}^{0}\Psi^{\dagger}\Psi=1~.
\end{equation}
In the case of 2 spatial dimensions, the Dirac equation (\ref{eqn:DiracCurved})
can be reduced, after some algebra, to
\begin{equation}
i\gamma^{a}\left[e_{a}^{\mu}\partial_{\mu}\Psi+\frac{1}{2\sqrt{|g|}}\partial_{\mu}(e_{a}^{\mu}\sqrt{|g|})\Psi\right]-m\Psi=0~,\label{eqn:Dirac2+1}
\end{equation}
where the $\gamma_{a}$ matrices become Pauli matrices. A simple
choice of the vierbeins obeying relation (\ref{eqn:VierbeinsDef})
is 
\begin{equation}
e_{0}=(e^{A(y)},0,0)\quad e_{1}=(0,e^{A(y)},0)\quad e_{2}=(0,0,1)~,
\end{equation}
which yields the following expression for the Dirac equation 
\begin{equation}
i\partial_{t}\Psi=-i\gamma^{0}\gamma^{1}\partial_{x}\Psi-i\gamma^{0}\gamma^{2}\partial_{y}(e^{-A(y)}\Psi)+\gamma_{0}e^{-A(y)}m\Psi~.\label{eqn:DiracExplicit}
\end{equation}
This expression can be rewritten in Hamiltonian form as 
\begin{equation}
i\partial_{t}\chi=\mathcal{H}\chi,\label{eqn:DiracSimp}
\end{equation}
with
\begin{equation}
\mathcal{H}=-\frac{i}{2}\{B^{x},\partial_{x}\}-\frac{i}{2}\{B^{y},\partial_{y}\}+\gamma_{0}e^{-A(y)}m~, \label{eq:HamiltonianDirac}
\end{equation}
where the change of variable $\chi=e^{-A(y)/2}\Psi$ was performed,
and we defined 
\begin{equation}
B^{x}=\gamma^{0}\gamma^{1}~,\quad B^{y}=e^{-A(y)}\gamma^{0}\gamma^{2}~.
\end{equation}
The symbol $\{\cdot,\cdot\}$ represents the anticommutator of two operators.
There is some freedom in the choice of the gamma matrices. For convenience,
we choose 
\begin{equation}
\gamma^{0}=\sigma_{x}~,\quad\gamma^{1}=i\sigma_{y}~,\quad\gamma^{2}=i\sigma_{z}~.
\end{equation}

\subsection*{Boundary Conditions for Fermionic Fields}

\label{sec:Boundary} The periodic condition (\ref{eqn:PeriodicCond})
simply implies that the fermionic fields need also to be periodic
\begin{equation}
\chi(x^{\mu},y+2L)=\chi(x^{\mu},y)~,\label{eqn:BoundaryPeriodic}
\end{equation}
but the $\mathbb{Z}_{2}$ needs a deeper consideration, since it has to leave
the fermionic action invariant. We can write the fermionic action
as 
\begin{equation}
S_{F}=  \int dx^{\mu}\int dy\overline{\chi}(x^{\mu},y)
 \left(i\gamma^{\mu}\partial_{\mu}+i\gamma^{2}\partial_{y}e^{-A(y)}-e^{-A(y)m}\right)\chi(x^{\mu},y)~.
\label{eqn:ActionFermion}
\end{equation}
which is extremized by the Dirac equation (\ref{eqn:DiracExplicit}).
Under the action of $\mathbb{Z}_{2}$ it becomes 
\begin{equation}
S_{F}=  \int dx^{\mu}\int dy\overline{\chi}(x^{\mu},-y)
 \left(i\gamma^{\mu}\partial_{\mu}-i\gamma^{2}\partial_{y}e^{-A(-y)}-e^{-A(-y)}m\right)\chi(x^{\mu},-y)~.
\label{eqn:ActionFermionZ2}
\end{equation}
We have to find an operator $M$, defined as $\chi(x^{\mu},-y)=M\chi(x^{\mu},y)$,
so as to allow the action to remain invariant. Action (\ref{eqn:ActionFermionZ2})
then becomes 
\begin{equation}
S_{F}=  \int dx^{\mu}\int dy\overline{\chi}(x^{\mu},y)\gamma^{0}M\gamma^{0}
  \left(i\gamma^{\mu}\partial_{\mu}-i\gamma^{2}\partial_{y}e^{-A(y)}-e^{-A(y)}m\right)M^{\dagger}\chi(x^{\mu},y)~,
\end{equation}
and establishes the following restrictions for $M$ to keep the action
(\ref{eqn:ActionFermion}) invariant, 
\begin{align}
 & \gamma^{0}M\gamma^{0}\gamma^{\mu}M^{\dagger}=\gamma^{\mu}~,\\
 & \gamma^{0}M\gamma^{0}\gamma^{2}M^{\dagger}=-\gamma^{2}~,\\
 & \gamma^{0}M\gamma^{0}M^{\dagger}=\mathbb{I}~,
\end{align}
where the first 2 conditions come from the kinetic terms of the action,
and the last one arises from the mass term. There does not exist a
solution for $M$ that solves all conditions simultaneously, although $M=\eta\sigma_{z}$
is a solution for the first 2, with $\eta=\pm1$. This means that
a constant mass term is forbidden. In the following we restrict ourselves
to the case where the ``bulk mass'' $m$ vanishes.
 The action of the fermionic field is therefore 
\begin{equation}
\begin{split}S_{F}= & \int dx^{\mu}dy\overline{\chi}(x^{\mu},y)\left(i\gamma^{\mu}\partial_{\mu}+i\gamma^{2}\partial_{y}e^{-A(y)}\right)\chi(x^{\mu},y)~,\end{split}
\label{eqn:ActionFermionNoMass}
\end{equation}
and the fermionic field has to obey the boundary condition 
\begin{equation}
\chi(x^{\mu},-y)=\eta\sigma_{z}\chi(x^{\mu},y)~,\label{eqn:BoundaryZ2}
\end{equation}
with $\eta=\pm1$.

\begin{figure*}
	\begin{minipage}{0.49\textwidth}
		\centering
		\includegraphics[width=\linewidth]{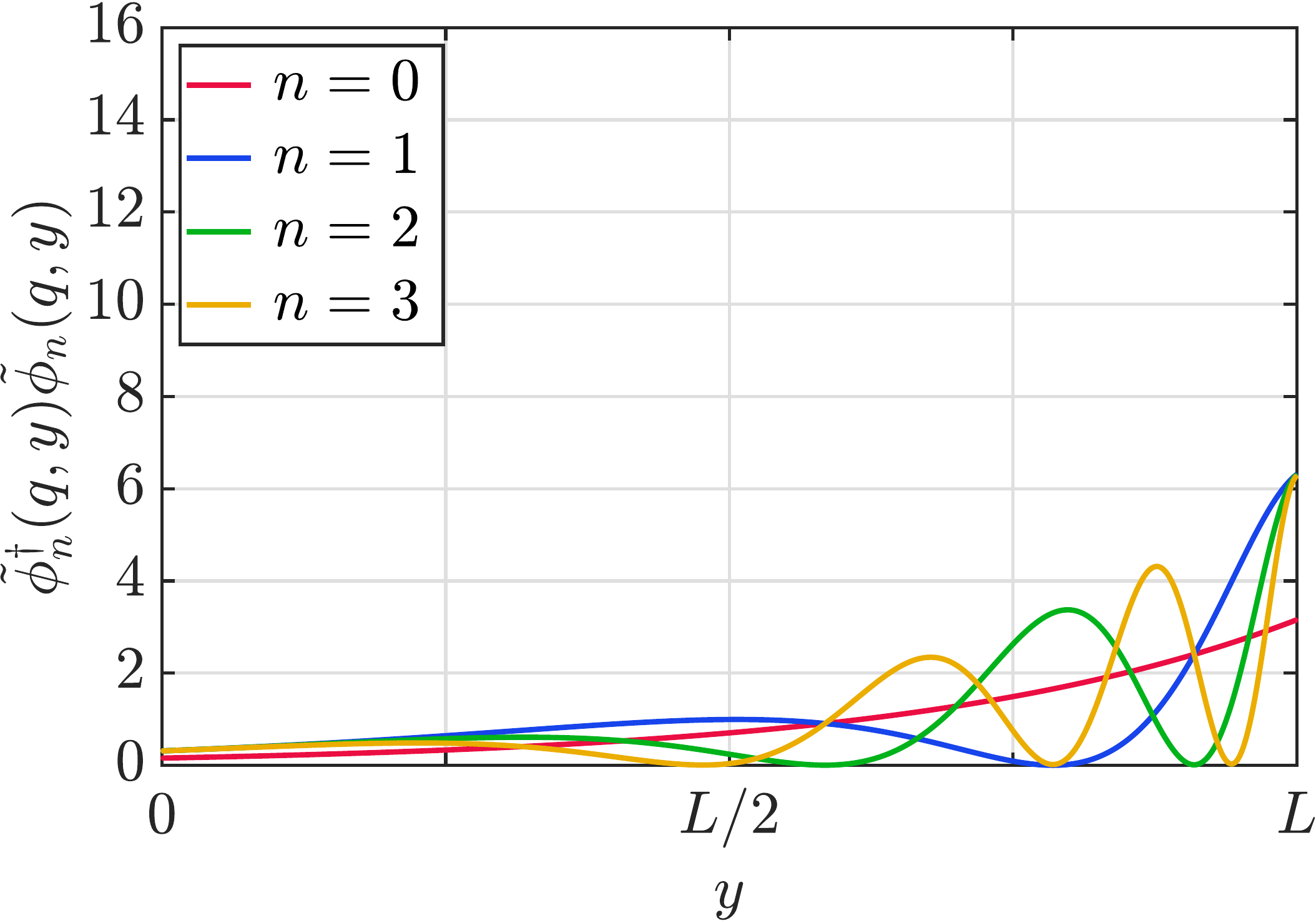}
	\end{minipage}
	\hfill
	\begin{minipage}{0.49\textwidth}
		\centering
		\includegraphics[width=\linewidth]{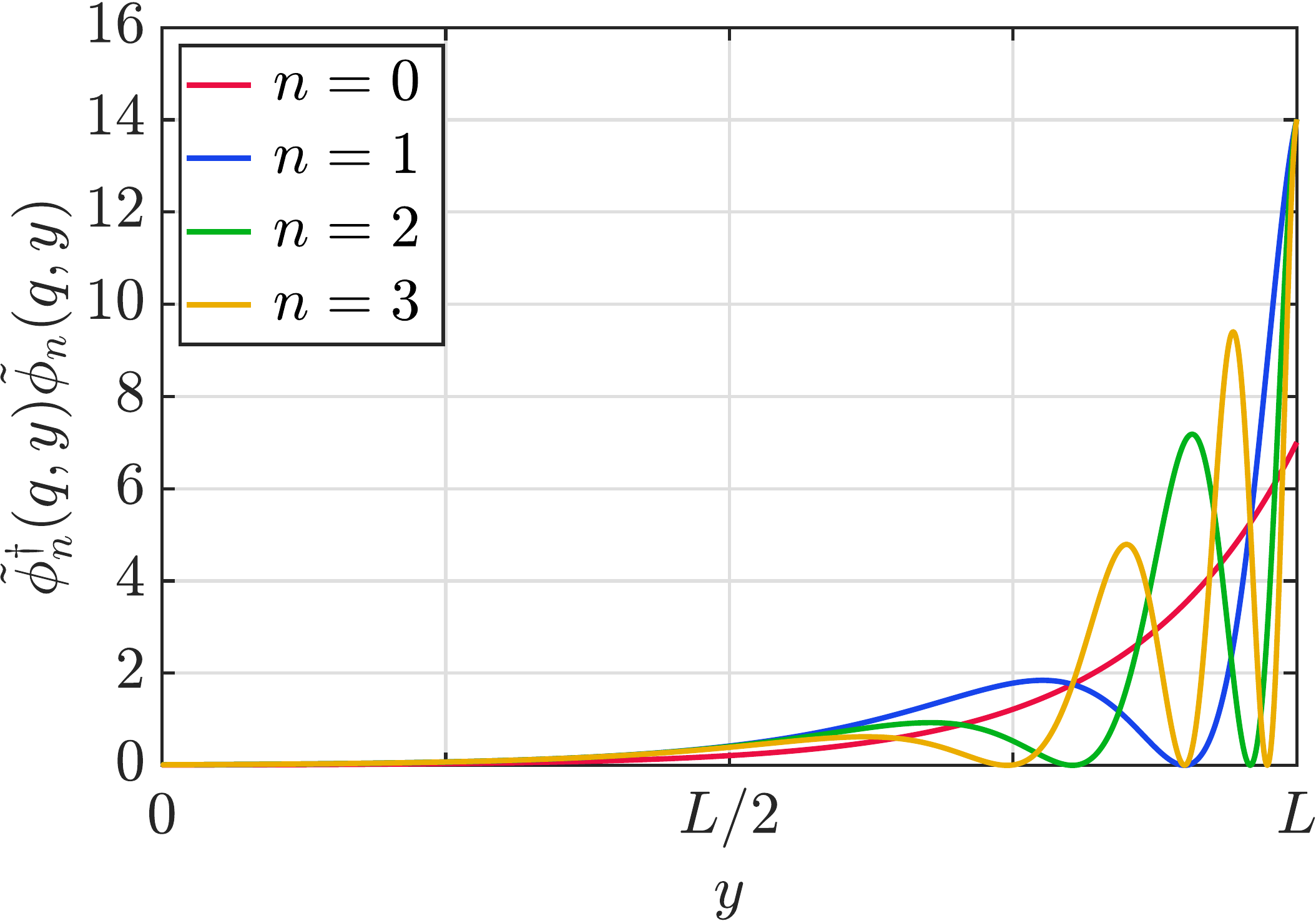}
	\end{minipage}
	\caption{Plots of the probability distribution for the first four stationary states, with positive
energy and a value of $q=10$, for $kL=3$ on the left and for $kL=7$ on the right.}
	\label{fig:Confinement} 
\end{figure*}

\subsection*{Stationary Solutions}
\label{sect:StationaryStates} In this model, the Dirac field satisfies
a complicated equation, Eq.~(\ref{eqn:DiracExplicit}), which is difficult
to address even numerically. In order to obtain some insight, we first
look for stationary solutions, which are defined as the eigenstates of the Hamiltonian.
For $m=0$, and with our choice of the gamma matrices, the Hamiltonian
takes the form 
\begin{equation}
\mathcal{H}=-\sigma_{z}\hat{p}_{x}+\frac{\sigma_{y}}{2}\left(e^{-A(y)}\hat{p}_{y}+\hat{p}_{y}e^{-A(y)}\right)~,\label{eqn:DiracMassless}
\end{equation}
where $p_{k}=-i\partial_{k}$ is the momentum operator along the $k$
direction ($k=x,y$). The stationary states $\phi_{n}(x,y)$ corresponding to
energy $E_{n}$ satisfy 
\begin{equation}
\mathcal{H}\phi_{n}(x,y)=E_{n}\phi_{n}(x,y)~.\label{eq:Eigenproblem}
\end{equation}
It is convenient to introduce a Fourier transform on the ordinary
dimension $x$: 
\begin{equation}
\tilde{\phi}_{n}(q,y)=\int dxe^{-iqx}\phi_{n}(x,y)~,
\end{equation}
since the field is free to move along this direction.  We found
the energies 
\begin{equation}
E_{n}=\pm\sqrt{q^{2}+\left(k\alpha_{n}\right)^{2}}~,\label{eqn:SpectrumSS}
\end{equation}
where 
\begin{equation}
\alpha_{n}=\frac{n\pi}{e^{kL}-1}~,~n=0,1,\;\dots
\end{equation}
The eigenfunctions associated with this spectrum that satisfy the
boundary condition (\ref{eqn:BoundaryZ2}), for the particular case
with $\eta=1$, are 
\begin{align}
\tilde{\phi}_{n}^{\uparrow}(q,y)&=\sqrt{\frac{2k}{e^{kL}-1}}  \frac{E_{n}+q}{\sqrt{(E_{n}+q)^{2}+(k\alpha_{n})^{2}}}
  e^{\frac{k|y|}{2}}\cos\left[\alpha_{n}\left(e^{k|y|}-1\right)\right]~,\label{eqn:EigenStateUpEta+}\\
\tilde{\phi}_{n}^{\downarrow}(q,y)&=\sqrt{\frac{2k}{e^{kL}-1}}  \frac{k\alpha_{n}}{\sqrt{(E_{n}+q)^{2}+(k\alpha_{n})^{2}}}
  e^{\frac{k|y|}{2}}\sin\left[\alpha_{n}\left(e^{k|y|}-1\right)\right]\mathrm{sign}(y)~,\label{eqn:EigenStateDownEta+}
\end{align}
where the components of the spinor field are $\tilde{\phi}_{n}=(\tilde{\phi}_{n}^{\uparrow},\tilde{\phi}_{n}^{\downarrow})^{T}$.
The particular case $n=0$ only has an upper component, which is given
by 
\begin{equation}
\tilde{\phi}_{0}^{\uparrow}(q,y)=\sqrt{\frac{k}{e^{kL}-1}}e^{\frac{k|y|}{2}}\mathrm{sign}(E_{n}+q)~,\label{eq:n0mode}
\end{equation}
and is undefined for energy and momentum with different sign. The procedure to obtain the eigenfunctions is detailed in the Supplementary Material, as well as the solution for $\eta = -1$. The
probability distribution associated to these wavefunctions is concentrated
around $y=L$ for high values of the warp coefficient $k$. We
illustrate this behavior in Fig.~\ref{fig:Confinement},
where we have plotted the probability density for the first modes
with positive energy, and momentum $q=10$, for a value of the warp
coefficient $kL=3$ and $kL=7$, respectively.

\section*{A Quantum Walk for the Randall-Sundrum model}

Once we have discussed the main properties of the RSM in the continuum
spacetime, we focus on the main goal of our work, which consists in
constructing a QW that is able to simulate the dynamics of a spin
$1/2$ particle subject to the geometric effects and symmetries of
the model. To incorporate the metric, we adapt the scheme introduced
in \cite{DebbaschWaves}, which allows to reproduce (in the continuum
limit) a Dirac equation of the form Eq.~(\ref{eqn:DiracSimp}).
The QW is defined on a 2-dimensional discrete grid with $x$ and $y$
axis, with discrete positions labeled by $r$ and $s$, respectively.
The grid points are equally spaced by $\epsilon$, so that the spatial
coordinates can be related to the grid points by $x=\epsilon r$ and
$y=\epsilon s$. The Hilbert space that corresponds to these spatial
degrees of freedom, $\mathcal{H}_{spatial}$ is spanned by the basis
$\{|x=\epsilon r,y=\epsilon s\rangle\}/r,s\in\mathbb{Z}$. Time steps
are labeled by $j\in\mathbb{Z}$, and are also equally spaced by $\epsilon$.
The coin (or internal) space is a 2 dimensional Hilbert space $\mathcal{H}_{\mathrm{coin}}$,
so that the total Hilbert space is $\mathcal{H}_{tot}=\mathcal{H}_{spatial}\otimes\mathcal{H}_{\mathrm{coin}}$.
At a given time step, the state of the walker will be represented
by a two component spinor $|\chi_{j}\rangle\in\mathcal{H}_{tot}$.
The one step evolution of the QW is given by 
\begin{equation}
|\chi_{j+1}\rangle=U|\chi_{j}\rangle~,\label{eqn:QWevolution}
\end{equation}
where $U$ is a unitary operator which consists on rotations in the
components of $|\chi_{j}\rangle$, and translations by $\epsilon$
in the two directions of physical space $x$ and $y$. The angles
of rotation can, in general, be dependent on the spacetime coordinates
of the walker. Following \cite{DebbaschWaves}, we adopt
\begin{equation}
U=R^{-1}(y)\left[\Theta(y)S_{y}(\epsilon/2)\right]^{2}R(y)S_{x}(\epsilon)~,\label{eq:Uoperator}
\end{equation}
where $S_{k}(\epsilon)=\exp(-i\sigma_{z}p_{k}\epsilon)$ are spin-dependent
shift operators in the direction $\pm k$ (with $k=x,y$),
\begin{equation}
\Theta(y)=\begin{pmatrix}c(y) & is(y)\\
-is(y) & c(y)
\end{pmatrix}~,\label{eq:Thetarotation}
\end{equation}
with $c(y)=e^{-A(y)}$, $s(y)=\sqrt{1-e^{-2A(y)}}$, and 
\begin{equation}
R(y)=\frac{1}{\sqrt{2}}\begin{pmatrix}f^{*}(y) & if(y)\\
-f^{*}(y) & -if(y)
\end{pmatrix}~,\label{eq:RRotation}
\end{equation}
where $f(y)=\sqrt{\frac{1+c(y)}{2}}+i\sqrt{\frac{1-c(y)}{2}}$. At
each position $(r,s)$ we introduce 
\begin{equation}
\chi_{j,r,s}\equiv\langle x=\epsilon r,y=\epsilon s|\chi_{j}\rangle=\begin{pmatrix}\chi_{j,r,s}^{\uparrow}\\
\chi_{j,r,s}^{\downarrow}
\end{pmatrix}~,\label{eq:Xicomponents}
\end{equation}
which represents the amplitude (given a component of the spin) for
the particle to be localized at the position labeled by $(r,s)$ and
time step $j$. In this way, the time step defined by (\ref{eqn:QWevolution})
can be recast as a recursive formula for $\chi_{j,r,s}$, which is
provided in the Supplementary Material. In order to implement this QW
to simulate fermions in the RSM, appropriate conditions have to be
set to comply with the boundary conditions \eqref{eqn:BoundaryPeriodic} and \eqref{eqn:BoundaryZ2}.
It can be explicitly shown, from the recursive formula for $\chi_{j,r,s}$, that this QW dynamics respects (\ref{eqn:BoundaryZ2}),
in the sense that, if the walker obeys the condition 
\begin{equation}
\chi_{j,r,-s}=\eta\sigma_{z}\chi_{j,r,s}~,\label{eqn:BoundaryQW}
\end{equation}
 at time $j$, it is also obeyed at time $j+1$. For the simulations,
we discretize the $y$ coordinate along the segment $[-L,L]$ with
a spacing $\epsilon$, and impose an initial condition which satisfies
Eq.~(\ref{eqn:BoundaryQW}). We use the same lattice spacing in the
$x$ direction, together with an strategy that adapts its effective
extension to the time step. We also impose periodic boundary conditions
on the grid to respect condition (\ref{eqn:BoundaryPeriodic}),
taking into account that functions evaluated at $y=L+\epsilon$ should
be identified with functions at $y=-L+\epsilon$ to respect the periodicity
in the range $[-L,L]$.

\section*{Results}

The QW defined in the previous section is guaranteed to reproduce (in
the continuum limit) a Dirac equation of the form \eqref{eqn:DiracSimp},
such as the one corresponding to the RSM. The question that arises
concerns the dynamics appearing at a finite lattice and time step
spacing. Of course, one does not expect the QW to behave exactly as
the continuum field but, to what extent do they differ?
Are there any new features that appear in the discrete case? In particular,
we are interested in looking for some kind of probability concentration
towards the visible brane, for a given initial condition. In this
Section we explore all these features.

\subsection*{Stationarity of the Eigenstates Solutions on the Quantum Walk}

\label{sec:StationaryQW} 
\begin{figure}[ht]
	\centering
	\includegraphics[width=0.4\linewidth]{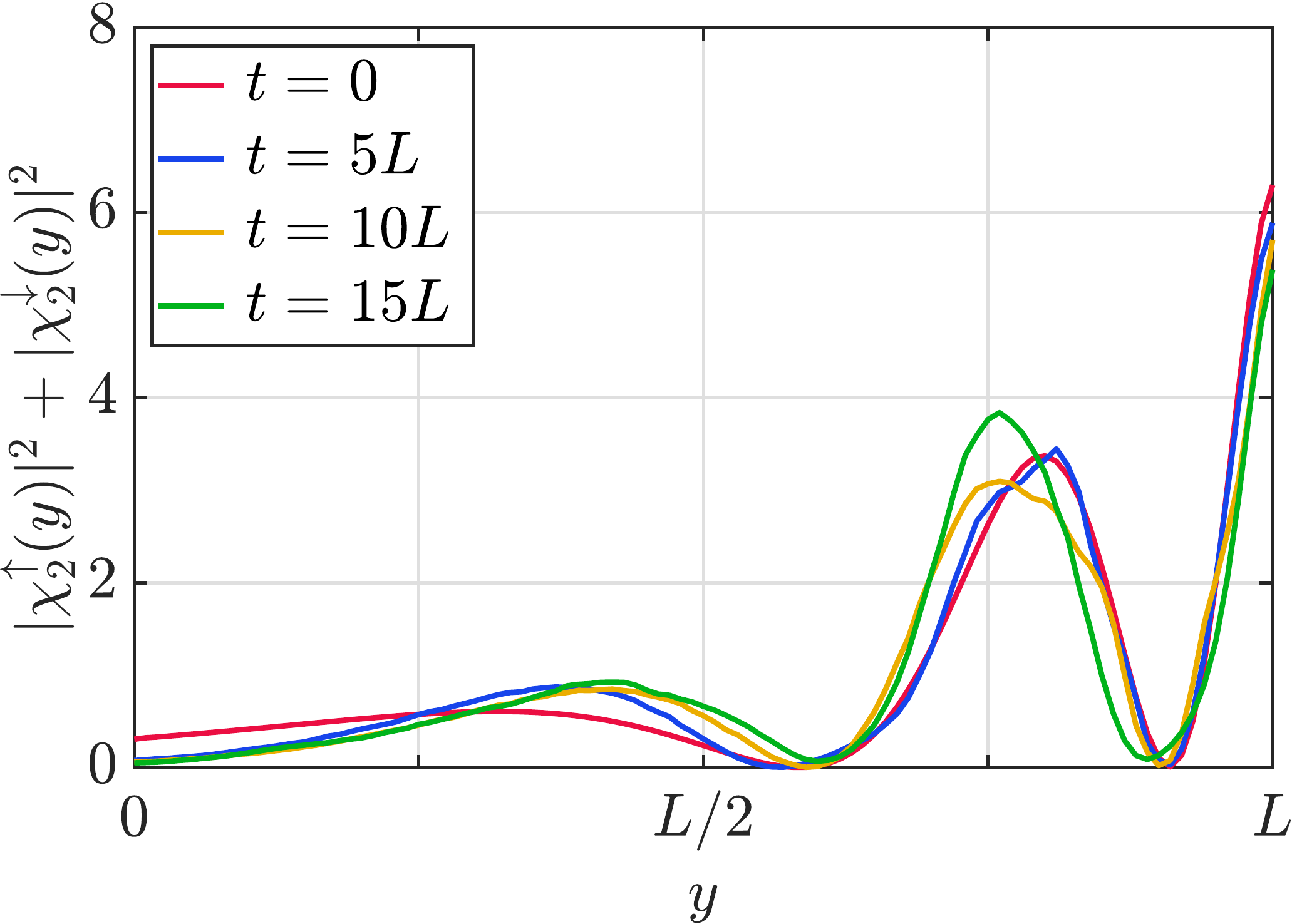}
	\caption{Snapshots of the probability density starting from an initial 
eigenstate with $n=2$ and positive energy, for a value of $kL=3$,
and $q=10$. The simulation grid has 100 points along the
$y$ direction, and enough points have been taken in the $x$ direction
to ensure that the total probability density does not leak outside the boundaries.}
	\label{fig:SS_evolution} 
\end{figure}

As an initial comparison, we start by considering
the discretized version of the eigenstates corresponding to
the continuum limit Hamiltonian, obtained before.
Such states remain stationary within this limit (i.e. they just evolve
by adopting a trivial phase). How do they evolve under the action of
the QW? We consider an initial state which corresponds to  an eigenstate of 
the continuum, with fixed momentum $q$, 
and check whether the QW evolution
of this state is stationary. The initial condition of the walker is therefore
\begin{equation}
\chi_{0,r,s}=\tilde{\phi}_{n}(q,\epsilon s)e^{iq\epsilon r}~,\label{eqn:InitialPlaneWave}
\end{equation}
which represents a constant probability density along the ordinary
dimension $x$. As expected, the QW evolution does not remain stationary,
although it keeps a close resemblance to the initial state. This can
be observed from Fig.~\ref{fig:SS_evolution}, where we represented
the normalized marginal probability along the $y$ direction of the
walker (after summing over $x$) at different time steps, for an initial
stationary state solution with $n=2$, and warp coefficient $kL=3$.

\subsection*{Localization in the QW\label{subsec:Confining-in-the}}

\begin{figure}[ht]
	\centering
	\includegraphics[width=0.8\linewidth,  trim=100 0 100 0]{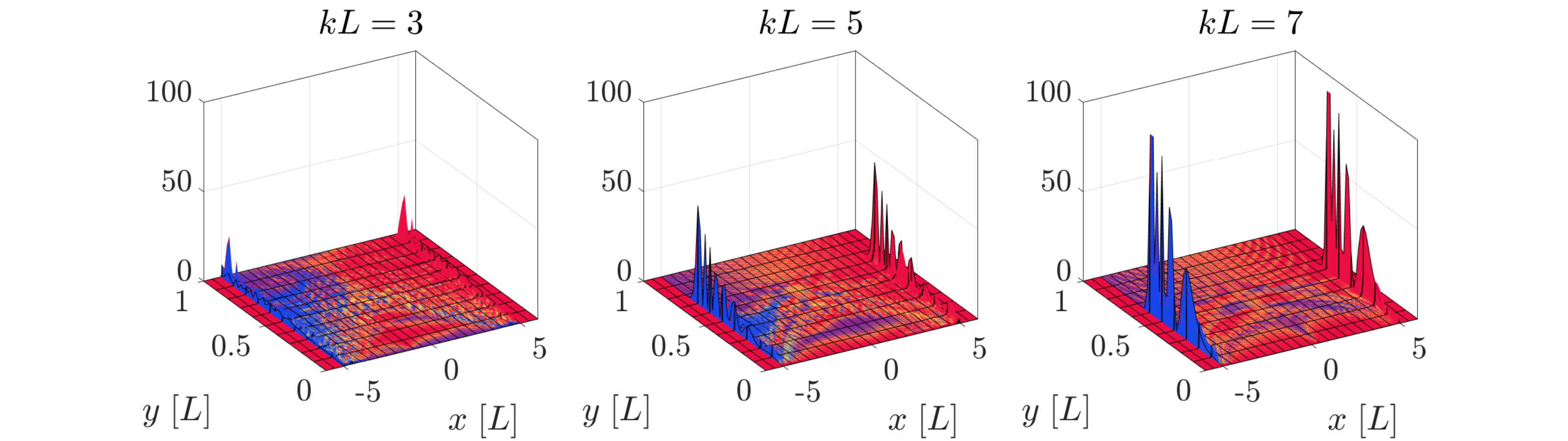}
	\caption{Probability density distribution, at $t=5 L$, of an initial localized walker centred at $(x_{0},y_{0})=(0,L/2)$ for different values of $kL$ with initial coin components $C_{0}=\frac{1}{\sqrt{2}}(1,i)^{T}$. The height of the curve represents the probability of finding the walker in that position, and the colors indicate the coin state. The red (blue) color indicates a predominance of the upper (lower) component, while yellow stands for a superposition of both components. The simulation grid has 100 points along the $y$ direction, and enough points have been taken in the $x$ direction to ensure that the total probability density does not leak outside the boundaries.}
	\label{fig:panell_confinament} 
\end{figure}

We now investigate the localization capability of the above defined
QW, i.e., whether it shows a tendency to concentrate the walker towards
the visible brane at $y=L$. We consider an initial walker which is
fully localized 
\begin{equation}
\chi_{0,r,s}^ {}=\delta_{x,0}\delta_{y,y_{0}}C_{0}~,\label{eqn:InitialDelta}
\end{equation}
where $C_{0}$ is the initial coin state, and we recall that $x=\epsilon r$
and $y=\epsilon s$. We explore the evolution of a walker which is
initially localized at the center of the extra dimension, that is
at $y_{0}=\frac{L}{2}$, and we study the probability distribution
for different values of the warp coefficient, at a given time step.
In Fig.~\ref{fig:panell_confinament} we show the surface plot of
the probability density with the above initial conditions, and $C_{0}=\frac{1}{\sqrt{2}}(1,i)^{T}$,
which induces a symmetric evolution in the ordinary dimension. The
blue (red) color of the surface represents dominance of the upper
(lower) coin component, while yellow stands for a superposition of
both components.

\begin{figure}[ht]
	\centering
	\includegraphics[width=0.4\linewidth]{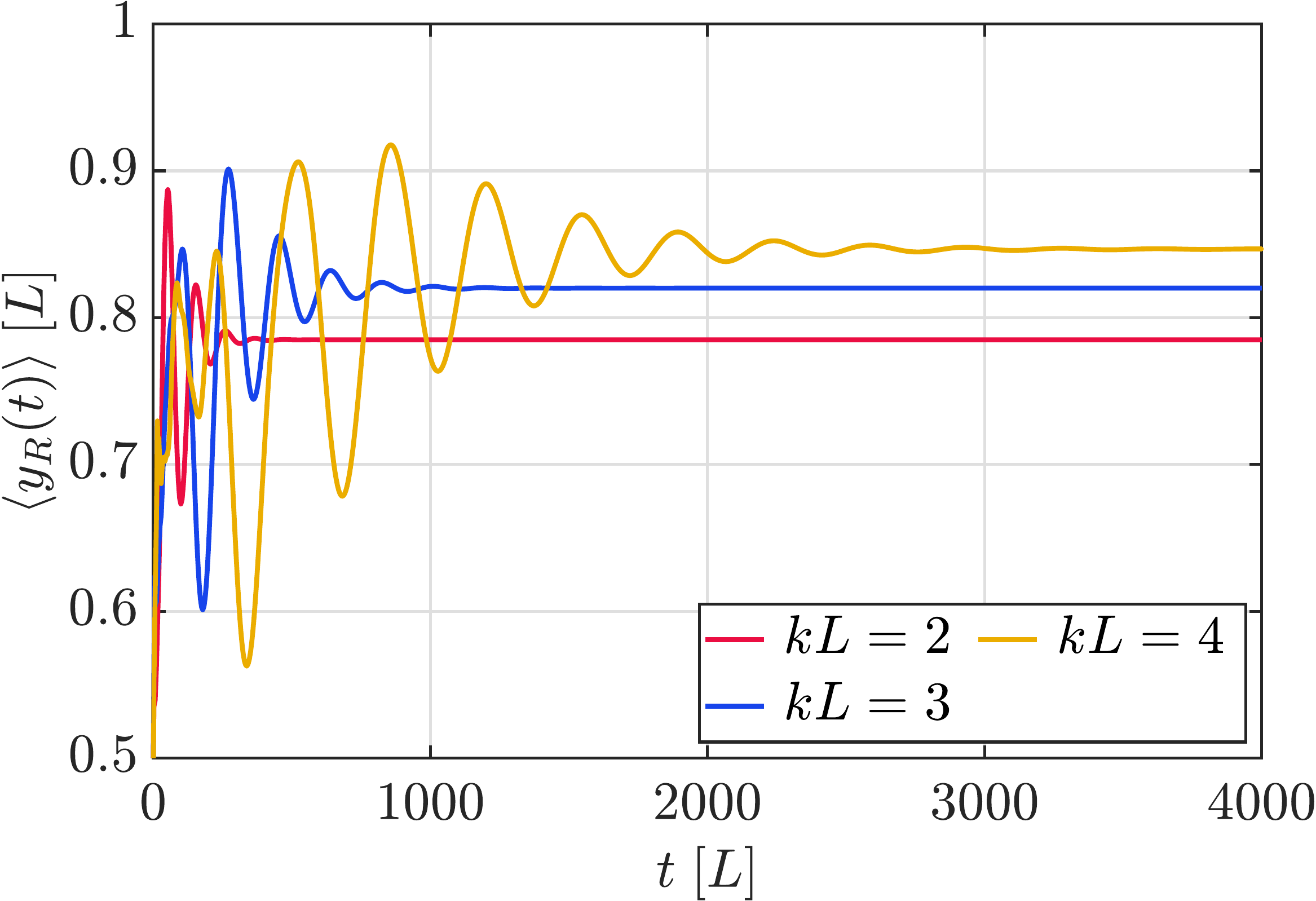}
	\caption{Expected value of the probability distribution along the extra dimension
$y$, as calculated from the FPD, for different values of the warp
coefficient $kL$. The initial condition is the same
as in Fig.~\ref{fig:panell_confinament}.}
	\label{fig:Expected value} 
\end{figure}

We notice that most of the probability distribution in the $x$ direction is concentrated along a freely propagating front which moves  at the maximum speed ($x=\pm t$), consistently with the fact that the QW simulates massless fermions.
We also notice that
most of the right propagating distribution (positive values of $x$)
is dominated by the upper coin component, while the part propagating
to the left (negative values of $x$) mainly contains the lower coin
component, a fact that can also be inferred from the explicit evolution
of the QW (see Supplementary Material for details). The propagation
of the walker along the extra dimension $y$ strongly depends on the
value of the warp coefficient. At $t=5L$, the distribution with the lowest value
of $kL$ possesses non-zero values on the visible brane $y=L$, while
the other two do not. In fact, the displacement of the probability
distribution towards $y=L$ is slower for the highest $kL$. In other
words, a larger value of the warp coefficient dramatically increases
the time scale of the dynamics along the extra dimension, and makes
it prohibitively expensive (in terms of computational cost) to explore
larger values of $kL$ than those considered here. 

In order to investigate whether the QW exhibits the same behavior
as the stationary states, in the sense that a higher value of the
warp coefficient induces a stronger localization near the visible
brane, we study the distribution of the freely propagated parts of
the walker (the regions around $x=\pm t$), where most of the probability
density is concentrated, as can be readily seen in Fig.~\ref{fig:panell_confinament}.
The probability distribution associated to these two zones will be
referred to as the ``freely propagating distribution'' (FPD). In
terms of the spinor components, those are the probability density
distributions obtained from $\chi_{j,s}^{\mathrm{R}}\equiv\chi_{j,j,s}$ and $\chi_{j,s}^{\mathrm{L}}\equiv\chi_{j,-j,s}$,
where $r=\pm j$ restricts the wavefunction to the two freely propagating
peaks. In Fig.~\ref{fig:Expected value} we represent the expected
value for these distributions along the $y$ dimension,
which can be defined as 
\begin{equation}
\langle y_{R(L)}(t)\rangle=\sum_{s}\epsilon s\;\chi_{j,s}^{\mathrm{R(L)}\dagger}\chi_{j,s}^{\mathrm{R(L)}}~,\label{eqn:ExpectedVal}
\end{equation}
where $t=\epsilon j$, for different values of $kL$. First of all
we notice that this quantity reaches an asymptotic value, which is closer
to $L$ for higher warp coefficients. Secondly, as discussed above, the warp
coefficient induces a change in the time scale of the dynamics, so
that lower values of the warp coefficient show a faster convergence
towards the asymptotic state, consistently with the features already observed in Fig.~\ref{fig:panell_confinament}.
\begin{figure}[h]
\hspace*{\fill}
	\begin{minipage}{0.4\linewidth}
		\centering
		\includegraphics[width=\linewidth]{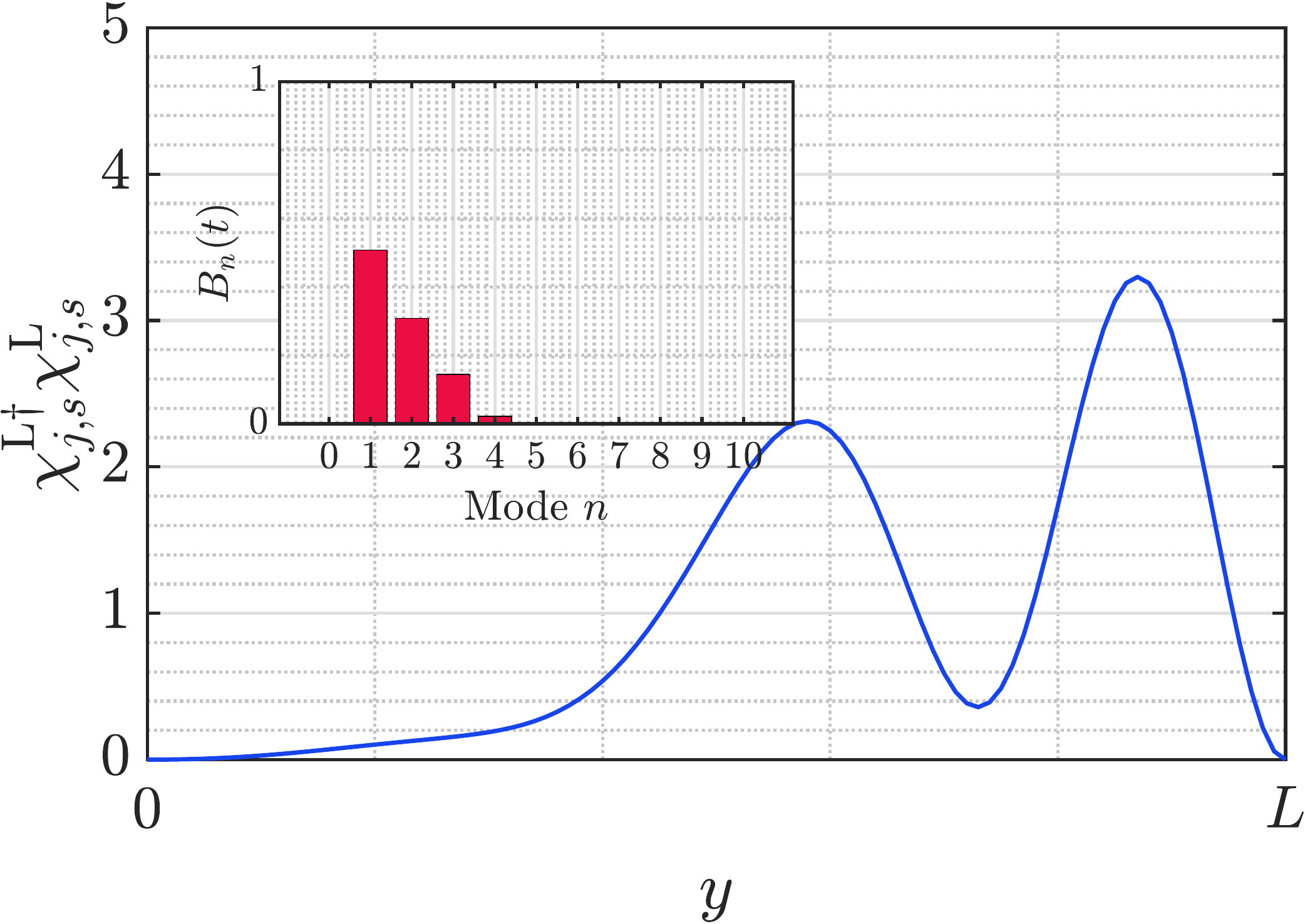}
	\end{minipage}
	\hspace*{0.05\linewidth}
	\begin{minipage}{0.4\linewidth}
		\centering
		\includegraphics[width=\linewidth]{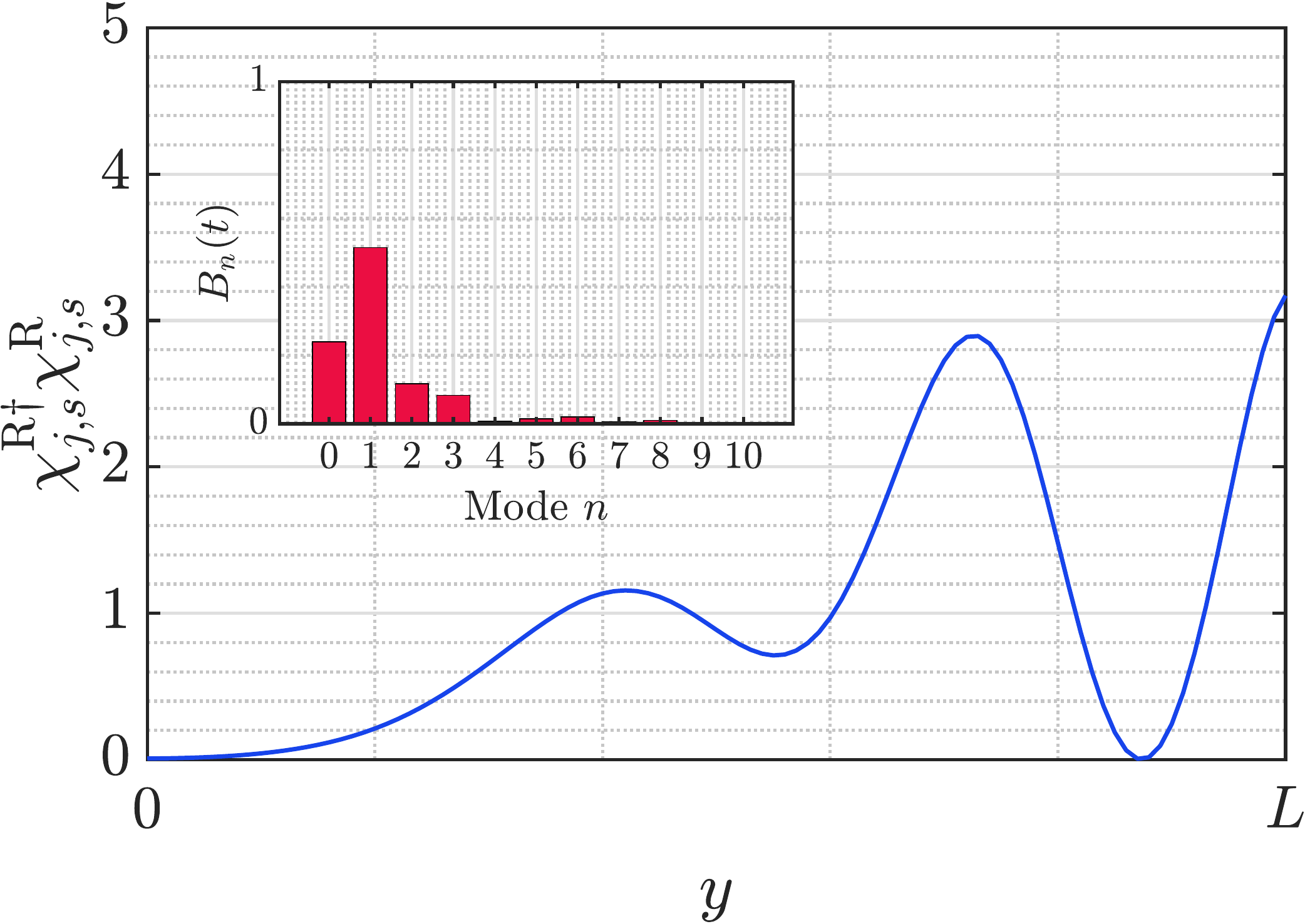}
	\end{minipage}
	\hspace*{\fill}
	\newline
	\hspace*{\fill}
		\begin{minipage}{0.4\linewidth}
		\centering
		\includegraphics[width=\linewidth]{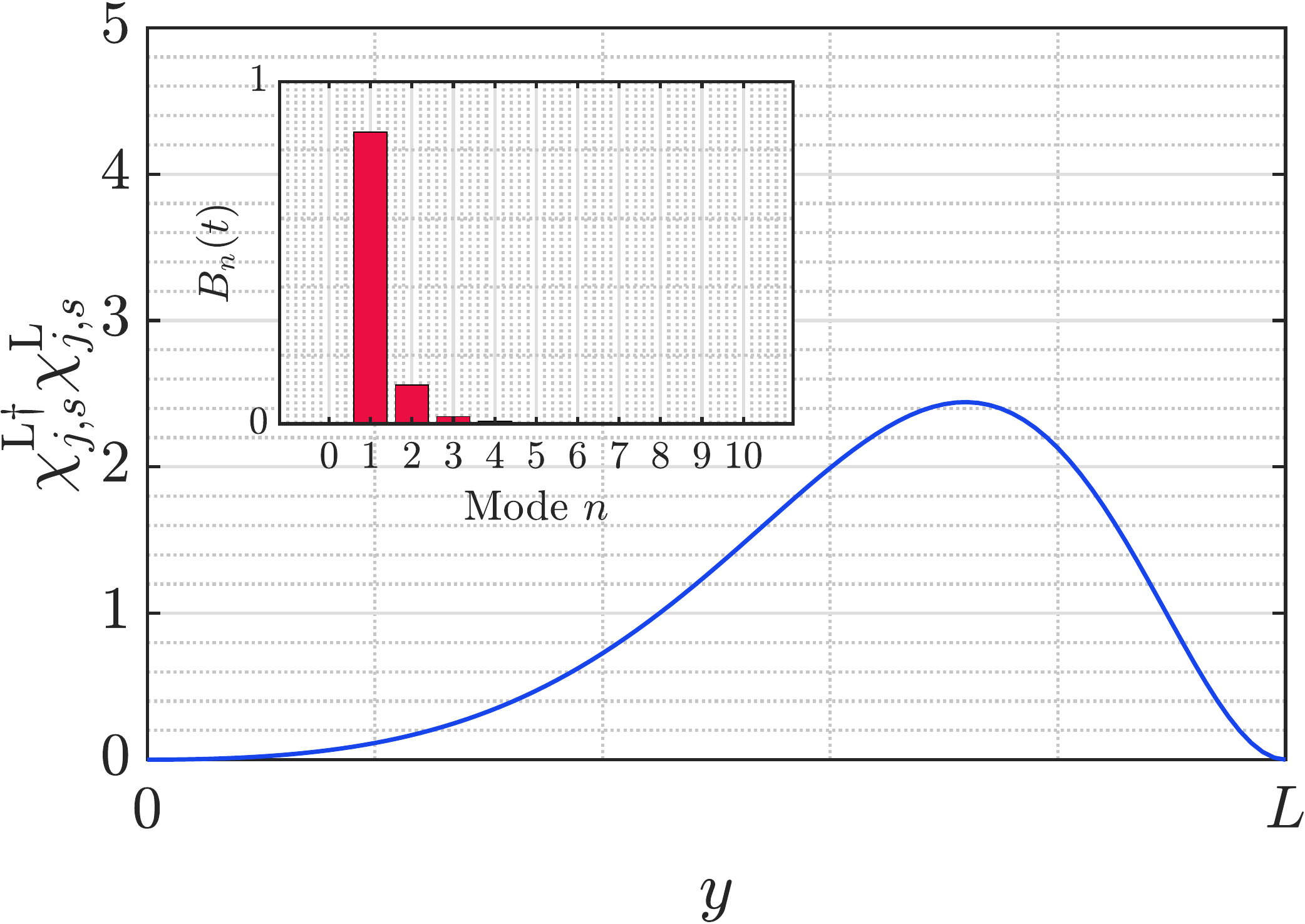}
	\end{minipage}
		\hspace*{0.05\linewidth}
	\begin{minipage}{0.4\linewidth}
		\centering
		\includegraphics[width=\linewidth]{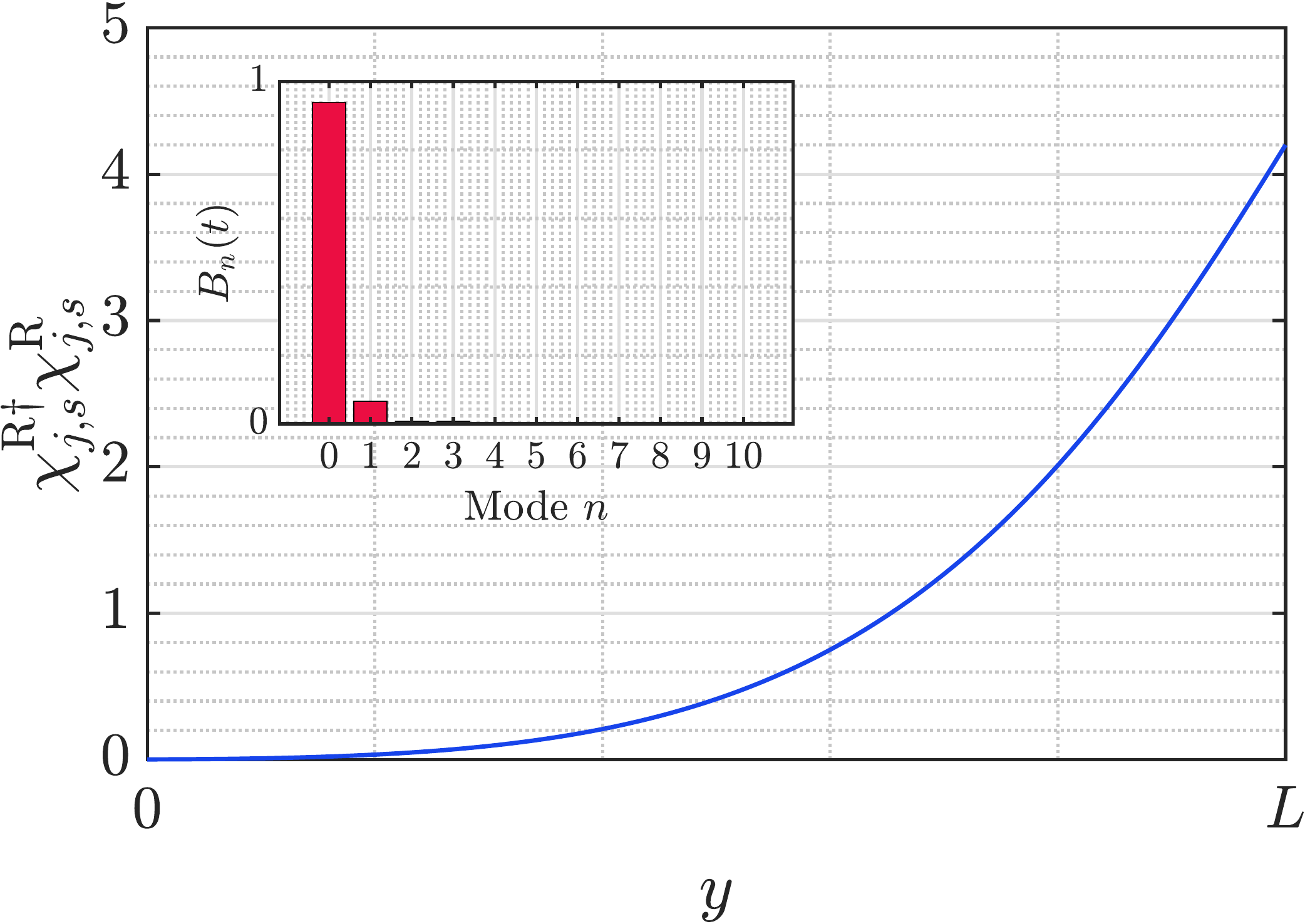}
	\end{minipage}
	\hspace*{\fill}
	\caption{Probability distributions of the FPDs along the extra dimension $y$,
for the value $kL=3$. The inset is a histogram showing
the value of the $B_{n}(t)$ coefficients, as defined by
Eq. (\ref{eqn:IntegratedModeCoeff}): see the text for an explanation. The left (right) panels show the left (right) FPD. The top panels are calculated at a shorter time $t=50 L$ and the bottom ones at a longer time $t=1000 L$.
The initial condition is the same as in Fig.~\ref{fig:Expected value},
and the simulation grid has 200 points along the $y$ direction.}
	\label{fig:ModeDecomposition}
\end{figure}

\subsection*{Mode decomposition of the freely propagating distribution}

Our simulations indicate that the FPD reaches a steady state along
the extra dimension, in a similar fashion as the expected value (\ref{eqn:ExpectedVal}). This evolution can be appreciated from the plots of Fig.~\ref{fig:ModeDecomposition}. Al late times (lower row), the probability distribution
resembles the probability density of a stationary state with positive
energy and momentum in one of the lowest modes: $n=0$ for the right
FPD, and $n=1$ for the left FPD. It is important to recall
that, as discussed above, the right
(left) FPD is predominantly composed by the upper (lower)
component of the spinor, and that $n=0$ has no lower component: see
Eq. (\ref{eq:n0mode}). This causes a fundamental difference when
comparing the left and right contributions. In order to investigate
these features on the time evolution, we introduce a decomposition on the wavefunction
of the walker as a combination of the stationary states basis. This allows us to write
\begin{equation}\label{eqn:ModeDecomposition}
\chi_{j,r,s}=\int_{-\pi/\epsilon}^{\pi/\epsilon}\frac{dq}{2\pi}\sum_{n}\beta_{n}(q,t)\tilde{\phi}_{n}(q,\epsilon s)e^{-iq\epsilon r}~,
\end{equation}
where the temporal dependence is included on the $\beta_{n}(q,t)$
coefficients. In the Supplementary Material we detail
how these factors can be computed, and define their normalization conditions.
In particular, we are interested on the contribution of each value
$n$, therefore we integrate out the dependence in the quasi-momentum
$q$. In other words, we are interested on the following (time-dependent)
coefficients: 
\begin{equation}
B_{n}(t)=\int_{-\pi/\epsilon}^{\pi/\epsilon}\frac{dq}{2\pi}\left|\beta_{n}(q,t)\right|^{2}~.\label{eqn:IntegratedModeCoeff}
\end{equation}
The different mode components $B_{n}(t)$ of Fig.~\ref{fig:ModeDecomposition} have been included as an inset in those plots.
On the one hand, it can be observed that, at long times, when a steady
state has been reached, the FPDs are mostly composed by the lowest possible
mode ($n=0$ or $n=1$, as discussed above). On the other hand, at
short times, the FPDs contain additional higher modes.

\subsection*{Entanglement Entropy \label{subsec:Entanglement-Entropy}}

\begin{figure}[h]
	\centering
	\includegraphics[width=0.5\linewidth]{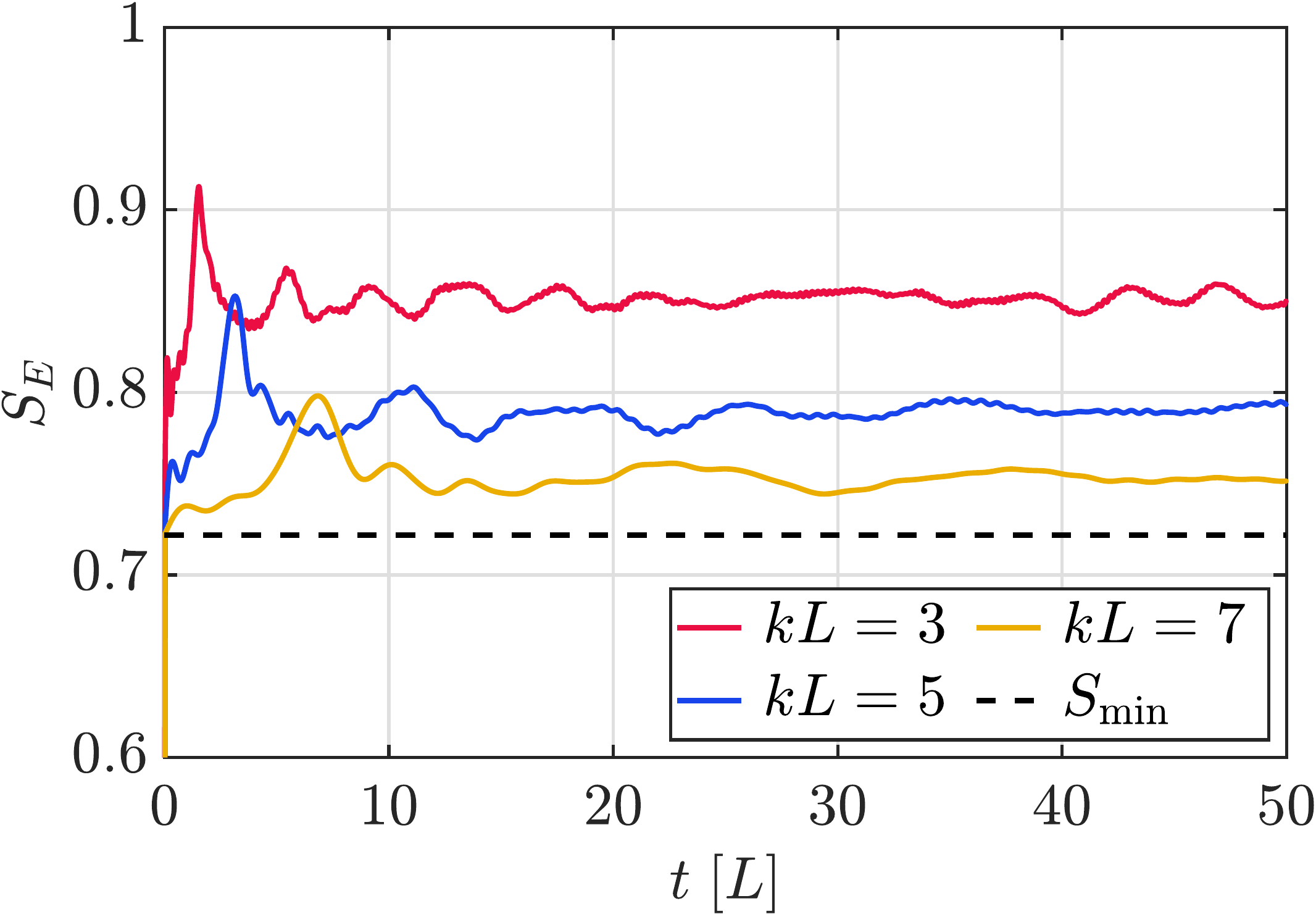}
	\caption{Evolution of the entanglement entropy with the initial condition
Eq.~(\ref{eqn:InitialDelta}) centred at $y_{0}=L/2$ for
different values of the warp coefficient and initial coin components
$C_{0}=\frac{1}{\sqrt{5}}(1,2i)$. The dotted line represent the minimum
value the entropy can reach for very high values of $kL$, which
is computed in the Supplementary Material. This simulation grid
has 50 points along the $y$ direction.}
	\label{fig:entropy} 
\end{figure}

\begin{figure}[h]
	\centering
	\includegraphics[width=0.8\linewidth, trim= 100 0 100 0]{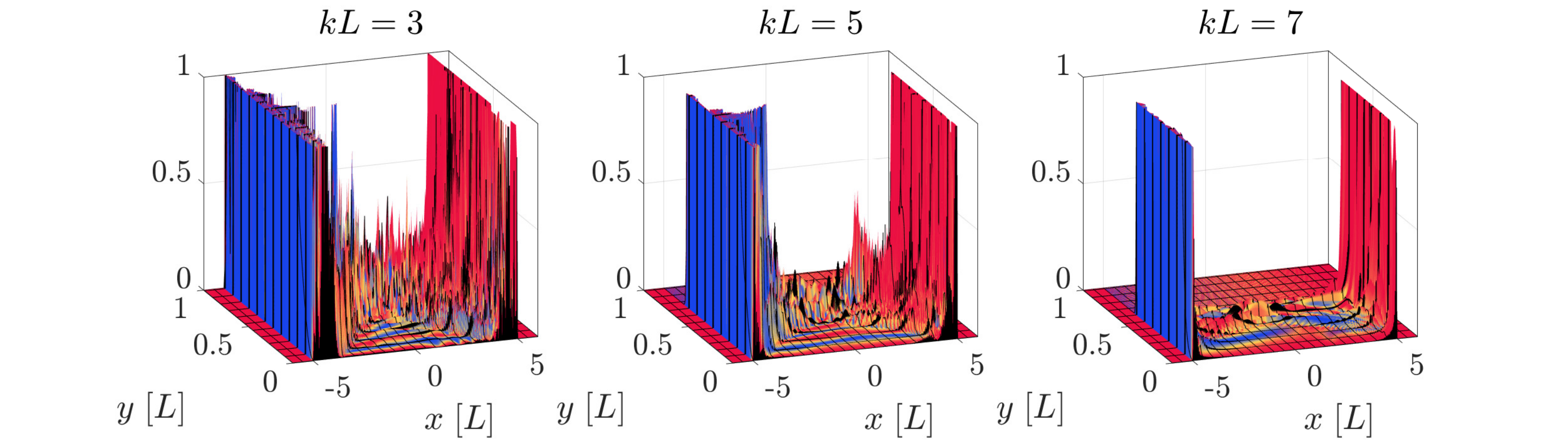}
	\caption{Probability density distribution, at $t=5 L$, for an initially localized
walker centered at $(x_{0},y_{0})=(0,L/2)$, and different values
of $kL$, with initial coin components $C_{0}=\frac{1}{\sqrt{5}}(1,2i)^{T}$.
The vertical axis has been zoomed in to show that the probability density
between the two regions of the FPD is more scrambled for lower values
of the warp coefficient. The colors and grid parameters are the same
as in Fig.~\ref{fig:panell_confinament}.}
	\label{fig:panell_confinament_zoom} 
\end{figure}

Finally, we study the entanglement properties that the QW exhibits
between the coin and position degrees of freedom for the already
considered, initially localized state. The entanglement
can be quantified using the von Neumann entropy of the reduced density
matrix in the coin space 
\begin{equation}
S(t)=-\mathrm{Tr}\left\{ \rho_{c}(t)\log_{2}\rho_{c}(t)\right\} ~,
\end{equation}
where $\rho_{c}(t=\epsilon j)=\sum_{r,s}\chi_{j,r,s}\chi_{j,r,s}^{\dagger}$
is the reduced density matrix in the coin space, i.e. after tracing
out the spatial degrees of freedom. In Fig.~\ref{fig:entropy} we
plot the evolution of the entanglement entropy of a fully localized
initial state for different values of the warp coefficient, with a
coin state $C_{0}=\frac{1}{\sqrt{5}}(1,2i)^{T}$. Notice that this
choice is different from that one used in the previous section,
for reasons that are explained below. It can be seen that the entanglement
entropy reaches lower values as $kL$ increases, an effect that can probably
be due to the fact that the probability density in between  the FPDs
becomes more spread (and therefore ``less ordered'') at lower values
of $kL$. This can be observed in Fig.~\ref{fig:panell_confinament_zoom},
where we plotted a zoomed version of Fig.~\ref{fig:panell_confinament},
but obtained with the above initial coin components $C_{0}=\frac{1}{\sqrt{5}}(1,2i)^{T}$.
One can see that, for lower values of the warp coefficient, a significant
part of the probability distribution is scrambled in the intermediate
region between both parts of the FPD. This diffusion effect can be
totally mitigated for extreme values of the warp coefficient, leading
to a minimum value of the entropy which is completely dominated by the FPD, and can be obtained from
the initial coin components. In the Supplementary Material we
show this limiting situation, and how the corresponding entropy can be
computed. The initial coin state $C_{0}=\frac{1}{\sqrt{2}}(1,i)^{T}$ previously used produces values of the entropy
which are very close to unity in all cases, making it difficult to
appreciate the effects that are discussed above.

\section*{Conclusions}

We have investigated a quantum walk which allows to simulate the Randall-Sundrum
model of extra dimensions, while satisfying the constrains imposed
by the symmetries of that model. This model has played an important
role in high energy physics, aiming to solve the hierarchy problem,
by introducing one finite extra dimension that possesses two branes
at its extremes. The matter fields are confined in the visible brane,
while gravity is allowed to span along this whole dimension. We worked
it out for the case of spin $1/2$ fermions in a two dimensional space, composed
by an ordinary dimension and an orbifolded one, apart from time, and
obtained the Dirac equation in this spacetime configuration. The boundary
conditions of the orbifold on the fermionic field forced it to be
massless on the bulk. In this lower dimensional space we were able to obtain the
eigenenergies of the fermionic field, as well as the corresponding
eigenstates, showing a probability density which is concentrated near
the visible brane, a phenomenon that bears an analogy with the localization
effect that can be found in many scenarios \cite{Lattices1956,aubry1980analyticity,PhysRevLett.49.833,Joye2010,PhysRevLett.106.180403,Navarrete-Benlloch2007,PhysRevE.82.031122}. 

This analogy motivated us to seek localization effects on the QW that
we introduced to simulate the RSM. The QW is defined in such a way that, in the
continuum limit, the Dirac equation of the fermionic
field for the RSM metric is recovered. We investigated the confining capabilities
of the QW, by considering an initially localized walker away from
the visible brane. We concluded that the freely propagating parts
of the probability distribution, where the probability is mostly concentrated,
reach an asymptotic value of the expected position along the extra
dimension. Moreover, the asymptotic value gets closer to the visible
brane for higher values of the warp coefficient, which therefore drives
the strength of localization, and also noticed that it had an effect
on the timescale of the dynamics, by delaying them for higher values
of the coefficient.

At long time steps, the probability densities show an asymptotic shape,
with a resemblance with the eigenstates that were obtained in the
continuous model, which suggested a study based on the decomposition
of the wavefunction in terms of these stationary states. We found
that the freely propagating parts of the QW are dominated, in the
asymptotic regime, by the lowest possible (i.e., compatible with the
symmetries of the model) modes. At intermediate time steps, the same
decomposition manifests a combination of multiple modes with higher
energy. 

Finally, we found that the entanglement between coin and spatial 
degrees of freedom
is reduced for stronger warp coefficients. We associated this result 
to the higher spreading of the density distribution for the lower values of the warp coefficient.

We conclude that quantum walks are suitable candidates for simulating
models of field theories with extra dimensions that rely on the curvature
of the spacetime. Not only the model is interesting from the point
of view of the field theory: It allows to design a quantum process
that can be tailored to exhibit very rich dynamics, showing free
propagation in one dimension, and an asymptotic confining behavior
on the other one, with rates that can be tuned by an appropriate choice
of the parameters. In this way, the interplay between high energy
physics and quantum simulations can be of mutual benefit.

\section*{Acknowledgements}

This work has been founded by the Spanish FEDER/MCIyU-AEI grant FPA2017-84543-P, SEV-2014-0398 and Generalitat Valenciana grant PROMETEO/2019/087.
We also acknowledge support from CSIC Research Platform PTI-001.

\bibliography{biblio}

\newpage
\section*{Supplementary Material}
\subsection*{Metric Solution}

\label{append:WrapFactor} Here we prove that the metric
(\ref{eqn:Metric}) extremizes the background action (\ref{eqn:ActionBackgroundTotal}),
which can be rewritten as 
\begin{equation}
S=\int dy\int dx^{\mu}\sqrt{|g|}\Big(2\alpha\mathcal{R}-\Lambda-\delta(y)T_{\mathrm{hid}}-\delta(y-L)T_{\mathrm{vis}}\Big)~.
\end{equation}
The extrema of this action gives the following Einstein equations
\begin{equation}
\sqrt{|g|}\bigg(R_{MN}  -\frac{1}{2}g_{MN}\mathcal{R}+\frac{1}{4\alpha}\Lambda g_{MN}\bigg)=
-\frac{\sqrt{|g|}}{4\alpha}  \bigg(T_{\mathrm{hid}}\delta(y)g_{\mu\nu}\delta_{M}^{\mu}\delta_{N}^{\nu}+T_{\mathrm{hid}}\delta(y-L)g_{\mu\nu}\delta_{M}^{\mu}\delta_{N}^{\nu}\bigg)~,
\end{equation}
with indices $M,N=\{t,x,y\}$, while $\mu,\nu=\{t,x\}$ only
account for ordinary dimensions. After computing the curvature tensor,
we obtain the following equation for the $yy$ component 
\begin{equation}
A'(y)^{2}+\frac{\Lambda}{4\alpha}=0~,\label{eqn:Einstein55}
\end{equation}
which yields the solution 
\begin{equation}
A(y)=k|y|~,\quad\mathrm{with}\quad k\equiv\sqrt{-\frac{\Lambda}{4\alpha}}~,\label{eqn:WrapFactorSol}
\end{equation}
and has to be consistent with the orbifold symmetry (\ref{eqn:Z2Cond}).
There are no $\mu y$ components, as the metric and tensors with these
components vanish. The $\mu\nu$ components of the Einstein equations
are 
\begin{equation}
(A'(y)^{2}-A''  (y))e^{-2A(y)}\eta_{\mu\nu}+\frac{1}{4\alpha}\Lambda e^{-2A(y)}\eta_{\mu\nu}=
-  \frac{1}{4\alpha}e^{-2A(y)}\eta_{\mu\nu}\left[T_{\mathrm{hid}}\delta(y)+T_{\mathrm{vis}}\delta(y-L)\right]~,
\end{equation}
which, making use of equation (\ref{eqn:Einstein55}), can be simplified
to 
\begin{equation}
A''(y)=\frac{1}{4\alpha}\left[T_{\mathrm{hid}}\delta(y)+T_{\mathrm{vis}}\delta(y-L)\right]~.\label{eqn:Einstein_munu}
\end{equation}
After computing the second derivative of $A(y)$ from Eq. (\ref{eqn:WrapFactorSol}),
and taking into account the periodicity of the metric (\ref{eqn:PeriodicCond}),
yields 
\begin{equation}
A''(y)=2k(\delta(y)-\delta(y-L))~,
\end{equation}
which allows us to identify, from Eq. (\ref{eqn:Einstein_munu}),
the values of the tensions 
\begin{equation}
-T_{\mathrm{vis}}=T_{\mathrm{hid}}=8\alpha k=\sqrt{-16\Lambda\alpha}~.
\end{equation}
The results obtained in this section indicate that the bulk geometry
has to be Anti-de Sitter, with a negative bulk cosmological constant,
and that the visible brane has negative tension, while the hidden
one is positive. These results differ, from standard works on Randall-Sundrum,
on the constant coefficient appearing in the expression for $k$,
Eq.~(\ref{eqn:WrapFactorSol}) because we are considering a one dimensional
ordinary space, so that the computations of the curvature tensor yield
different constant factors.

\subsection*{Hamiltonian eigenstates}

\label{append:SteadyState} In order to solve the eigenvalue problem
\eqref{eq:Eigenproblem}, it is convenient to perform the change of
basis $\xi(q,y)=H\tilde{\phi}(q,y)$, with 
\begin{equation}
	H=\begin{pmatrix}
	1 & 1 \\
	1 & -1
	\end{pmatrix}
\end{equation}
the Hadamard matrix,
so that the eigenvalue equation becomes 
\begin{align}
\left[q+\frac{i}{2}\left(p_{y}e^{-A(y)}+e^{-A(y)}p_{y}\right)\right]\xi_{n}^{-}=E_{n}\xi_{n}^{+}~,\label{eqn:Differential1stOrd-Pos}\\
\left[q-\frac{i}{2}\left(p_{y}e^{-A(y)}+e^{-A(y)}p_{y}\right)\right]\xi_{n}^{+}=E_{n}\xi_{n}^{-}~,\label{eqn:Differential1stOrd-Neg}
\end{align}
where $\xi^{\pm}$ are the components of $\xi=(\xi^{+},\xi^{-})^{T}$.
This system of equations can be decoupled, giving
\begin{equation}
\left[q^{2}+\frac{1}{4}\left(p_{y}e^{-A(y)}+e^{-A(y)}p_{y}\right)^{2}\right]\xi_{n}^{\pm}=E_{n}^{2}\xi_{n}^{\pm}~,\label{eqn:Differential2ndOrd}
\end{equation}
which is a second order differential equation that can be solved for
the appropriate boundary conditions. We solve this equation both in
the positive $\left[\xi^{\pm}(0<y<L)\right]_{P}$ and negative domain
$\left[\xi^{\pm}(-L<y<0)\right]_{N}$, delivering 
\begin{align}
\left[\xi_{n}^{\pm}(y)\right]_{P}= & Ae^{\frac{ky}{2}}\cos\left(e^{ky}\alpha_{n}\right)+Be^{\frac{ky}{2}}\sin\left(e^{ky}\alpha_{n}\right)~,\label{eqn:GenericSolution}\\
\left[\xi_{n}^{\pm}(y)\right]_{N}= & Ce^{-\frac{ky}{2}}\cos\left(e^{-ky}\alpha_{n}\right)+De^{\frac{ky}{2}}\sin\left(e^{-ky}\alpha_{n}\right)~,
\end{align}
where we defined 
\begin{equation}
\alpha_{n}=\frac{\sqrt{E_{n}^{2}-q^{2}}}{k}~.
\end{equation}
These solutions are related by the continuity conditions 
\begin{equation}
\left[\xi_{n}^{\pm}(0)\right]_{P}=\left[\xi_{n}^{\pm}(0)\right]_{N}~,\;\text{and}\;\left[\xi_{n}^{\pm}(L)\right]_{P}=\left[\xi_{n}^{\pm}(-L)\right]_{N}~,
\end{equation}
where in the last one the periodicity of the wavefunctions, Eq.~(\ref{eqn:BoundaryPeriodic}),
has been used, and imply that the solutions are related by $A=C$
and $B=D$. The discontinuity introduced by the delta terms at $y=0$
and $y=\pm L$, coming from $A''(y)$, imposes $B=A\tan\alpha_{n}$,
and the following restrictions to the energies 
\begin{equation}
\tan\alpha_{n}=\tan\left(e^{kL}\alpha_{n}\right)~,
\end{equation}
which yields the spectrum in Eq.~(\ref{eqn:SpectrumSS}). After taking
into account these conditions, the eigenstates become 
\begin{equation}
\xi_{n}^{\pm}(y)=Ae^{\frac{k|y|}{2}}\left[\cos\left(e^{k|y|}\alpha_{n}\right)+\tan\alpha_{n}\sin\left(e^{k|y|}\alpha_{n}\right)\right]~.
\end{equation}
However, these solutions come from the second order differential equation
(\ref{eqn:Differential2ndOrd}), whereas the original equations were
first order, and relate $\xi^{+}(y)$ to $\xi^{-}(y)$. To find the
appropriate solution of the eigenfunctions, we need to take into account
these relations. Since any lineal combination of solutions is also a
solution of the equations, we consider the solution $\left[\xi_{n}^{+}(y)\right]_{2}$
of Eq.~(\ref{eqn:Differential2ndOrd}) to obtain $\left[\xi_{n}^{-}(y)\right]_{1}$
from Eq.~(\ref{eqn:Differential1stOrd-Neg}), where $[~\cdot~]_{i}$
denotes whether the solution comes from a first ($i=1$) or second
($i=2$) order differential equation. Similarly, from $\left[\xi_{n}^{-}(y)\right]_{2}$
we obtain $\left[\xi_{n}^{+}(y)\right]_{1}$, so that 
\begin{equation}
\left[\xi_{n}^{\pm}(y)\right]_{1}=\bigg[\sin\left(\alpha_{n}e^{k|y|}\right)\left(\frac{q}{E}\tan\alpha_{n}\pm\frac{k\alpha_{n}}{E}\mathrm{sign}(y)\right)+
\cos\left(\alpha_{n}e^{k|y|}\right)\left(\frac{q}{E}\mp\frac{k\alpha_{n}}{E}\tan\alpha_{n}\mathrm{sign}(y)\right)\bigg]Ae^{\frac{k|y|}{2}}~.
\end{equation}
The general solution for the eigenstates is a lineal combination of
this pair of solutions 
\begin{align}
\xi_{n}^{+}(y)=K_{1}\left[\xi_{n}^{+}(y)\right]_{2}+K_{2}\left[\xi_{n}^{+}(y)\right]_{1}~,\\
\xi_{n}^{-}(y)=K_{1}\left[\xi_{n}^{-}(y)\right]_{1}+K_{2}\left[\xi_{n}^{-}(y)\right]_{2}~,
\end{align}
where the relation between the constants $K_{1}$ and $K_{2}$ is
set by Eq.~(\ref{eqn:BoundaryZ2}), which, depending on the possible
values of $\eta$, implies the restrictions 
\begin{align}
\eta=+1 & \implies K_{1}=K_{2}~,\\
\eta=-1 & \implies K_{1}=-K_{2}~.
\end{align}
Finally, undoing the change of basis, we recover the original eigenstate
components of Eqs.~(\ref{eqn:EigenStateUpEta+},\ref{eqn:EigenStateDownEta+})
for $\eta=+1$, while 
\begin{align}
\phi_{n}^{\uparrow}(y)&=\sqrt{\frac{2k}{e^{kL}-1}}  \frac{k\alpha_{n}}{\sqrt{(E_{n}+q)^{2}+(k\alpha_{n})^{2}}}
  e^{\frac{k|y|}{2}}\sin\left[\alpha_{n}\left(1-e^{k|y|}\right)\right]\mathrm{sign}(y)~,\label{eqn:EigenStateUpEta-}\\
\phi_{n}^{\downarrow}(y)&=\sqrt{\frac{2k}{e^{kL}-1}}  \frac{E_{n}+q}{\sqrt{(E_{n}+q)^{2}+(k\alpha_{n})^{2}}}
  e^{\frac{k|y|}{2}}\cos\left[\alpha_{n}\left(1-e^{k|y|}\right)\right]~,\label{eqn:EigenStateDownEta-}
 \end{align}
are obtained for $\eta=-1$, and where the remaining constant was set
by the normalization of the wavefunction 
\begin{equation}
\int_{0}^{L}dy\tilde{\phi}_{n}(q,y)^{\dagger}\tilde{\phi}_{n}(q,y)=1~.\label{eqn:SS_Norm_Cont}
\end{equation}
The solution for the particular case of $n=0$ has only a lower component,
and is given by 
\begin{equation}
\phi_{0}^{\downarrow}(y)=\sqrt{\frac{k}{e^{kL}-1}}e^{\frac{k|y|}{2}}\mathrm{sign}(E_{n}+q)~.
\end{equation}

\subsection*{QW explicit time step}

\label{append:QW} Making use of the equations that define the QW,
Eqs. (\ref{eqn:QWevolution}, \ref{eq:Uoperator}, \ref{eq:Thetarotation})
and (\ref{eq:RRotation}), one can recast the evolution of $|\chi_{j}\rangle$
as a recurrence relation relating the spinor components Eq. (\ref{eq:Xicomponents})
at two consecutive time steps. We arrive at 
\begin{align}
\chi_{j+1,r,s}^{\uparrow}= & -\frac{i}{2}e^{i\theta(y)}\left[s\left(y+\frac{\epsilon}{2}\right)+s\left(y-\frac{\epsilon}{2}\right)\right]\chi_{j,r+1,s}^{\uparrow}
 -\frac{1}{2}\left[s\left(y+\frac{\epsilon}{2}\right)-s\left(y-\frac{\epsilon}{2}\right)\right]\chi_{j,r-1,s}^{\downarrow}\nonumber \\
 & +\frac{1}{2}f(y)f(y+\epsilon)c\left(y+\frac{\epsilon}{2}\right)\chi_{j,r+1,s+1}^{\uparrow}
  +\frac{1}{2}f(y)f(y-\epsilon)c\left(y-\frac{\epsilon}{2}\right)\chi_{j,r+1,s-1}^{\uparrow}\nonumber \\
 & +\frac{i}{2}f(y)f^{*}(y+\epsilon)c\left(y+\frac{\epsilon}{2}\right)\chi_{j,r-1,s+1}^{\downarrow}
  -\frac{i}{2}f(y)f^{*}(y-\epsilon)c\left(y-\frac{\epsilon}{2}\right)\chi_{j,r-1,s-1}^{\downarrow}~,\label{eq:mapup}
\end{align}
for the upper component, where we recall that $y=\epsilon s$, and
we defined $e^{\pm i\theta(y)}=c(y)\pm is(y)$. For the lower component
one finds
\begin{align}
\chi_{j+1,r,s}^{\downarrow}= & \frac{i}{2}e^{-i\theta(y)}\left[s\left(y+\frac{\epsilon}{2}\right)+s\left(y-\frac{\epsilon}{2}\right)\right]\chi_{j,r-1,s}^{\downarrow}
 -\frac{1}{2}\left[s\left(y+\frac{\epsilon}{2}\right)-s\left(y-\frac{\epsilon}{2}\right)\right]\chi_{j,r+1,s}^{\uparrow}\nonumber \\
 & +\frac{1}{2}f^{*}(y)f^{*}(y+\epsilon)c\left(y+\frac{\epsilon}{2}\right)\chi_{j,r-1,s+1}^{\downarrow}
  +\frac{1}{2}f^{*}(y)f^{*}(y-\epsilon)c\left(y-\frac{\epsilon}{2}\right)\chi_{j,r-1,s-1}^{\downarrow}\nonumber \\
 & -\frac{i}{2}f^{*}(y)f(y+\epsilon)c\left(y+\frac{\epsilon}{2}\right)\chi_{j,r+1,s+1}^{\uparrow}
  +\frac{i}{2}f^{*}(y)f(y-\epsilon)c\left(y-\frac{\epsilon}{2}\right)\chi_{j,r+1,s-1}^{\uparrow}~.\label{eq:mapdown}
\end{align}
We notice that the upper components are displaced in one direction along the $x$ dimension, while the lower components are displaced in the opposite direction.

\subsection*{Mode decomposition of the freely propagating distribution}

\label{append:ModeDecomposition}

The stationary states found above
form an orthonormal basis, in the continuum limit, that allow for
a decomposition of any function along the $y$ coordinate, for a given
value of $q$. They can also be used, after a proprer discretization, 
in the lattice on which the QW is defined. Following this idea, we
introduced the decomposition in Eq.~\eqref{eqn:ModeDecomposition},
which is a function in the space of $q$, the lattice quasimomentum along the $x$ coordinate.
For this quasimomentum space, the spinor components are related to
Eq. (\ref{eq:Xicomponents}) via a discrete Fourier transform 
\begin{equation}
\tilde{\chi}_{j,s}(q)=\sum_{r}e^{-iq\epsilon r}\chi_{j,r,s}~.
\end{equation}
Making use of 
\begin{equation}
\sum_{r}e^{ix(q-q')}=\frac{2\pi}{\epsilon}\delta(q-q')~,
\end{equation}
and the orthonormality condition (\ref{eqn:SS_Norm_Cont}) on the
grid 
\begin{equation}
\epsilon\sum_{s}\tilde{\phi}_{n}(q,\epsilon s)^{\dagger}\tilde{\phi}_{m}(q,\epsilon s)=\delta_{n,m}~,
\end{equation}
the coefficients can be obtained as 
\begin{equation}
\beta_{n}(q,t)=\epsilon^{2}\sum_{s}\tilde{\phi}_{n}(q,\epsilon s)\tilde{\chi}_{j,s}(q)~.
\end{equation}
The coefficients of the freely propagating distribution with $x=t$
are 
\begin{equation}
\beta_{n}(q,t)=\epsilon^{2}\sum_{s}\tilde{\phi}_{n}(q,\epsilon s)e^{-iqt}\chi_{j,j,s}~,
\end{equation}
while, for $x=-t$, they read as 
\begin{equation}
\beta_{n}(q,t)=\epsilon^{2}\sum_{s}\tilde{\phi}_{n}(q,\epsilon s)e^{iqt}\chi_{j,-j,s}~.
\end{equation}
From the normalization condition of the spinor on the grid 
\begin{equation}
\epsilon^{2}\sum_{r,s}\chi_{j,r,s}^{\dagger}\chi_{j,r,s}=1~,
\end{equation}
and making use of the definition \eqref{eqn:ModeDecomposition}, it can be shown that the mode coefficients satisfy 
\begin{equation}
\sum_{n}\int_{-\pi/\epsilon}^{\pi/\epsilon}\frac{dq}{2\pi}\left|\beta_{n}(q,t)\right|^{2}=1~,
\end{equation}
which can be expressed in terms of the integrated coefficients (\ref{eqn:IntegratedModeCoeff})
as 
\begin{equation}
\sum_{n}B_{n}(t)=1~.
\end{equation}

\subsection*{High $kL$ limit of the QW time step and limiting entropy}

\label{append:QWlimitK} In the limit of a high warp factor $kL$,
the exponential $e^{-A(L)}$ becomes very small, so that the QW discrete
time recursive evolution Eqs. (\ref{eq:mapup},\ref{eq:mapdown})
can be expanded up to the lowest order in this factor, giving 
\begin{align}
\chi_{j+1,r,s}^{\uparrow}= & \chi_{j,r+1,s}^{\uparrow}~,\nonumber \\
\chi_{j+1,r,s}^{\downarrow}= & \chi_{j,r-1,s}^{\downarrow} \label{eqn:QWLimit}~.
\end{align}
Although this expansion is only valid for values
of $y$ close to $L$, it is still accurate enough for the initial
condition located at $y=L/2$. As discussed in the main text,
the asymptotic value of the entanglement entropy decreases as $kL$
is increased. Therefore, the minimum value of the entropy is reached
in the limit $e^{-A(L)}\approx0$. The initial condition $\chi_{0,r,s} = \delta_{r,0}\delta_{s,s_0}C_0$
can be iterated with the help of Eqs.~\eqref{eqn:QWLimit} to produce the explicit time evolution
\begin{align}
\chi_{j,r,s}^{\uparrow}= & C_{0}^{\uparrow}\delta_{r,j}\delta_{s,s_0}~,\nonumber \\
\chi_{j,r,s}^{\downarrow}= & C_{0}^{\downarrow}\delta_{r,-j}\delta_{s,s_0}~.
\end{align}
The corresponding reduced density matrix becomes time-independent
and diagonal: 
\begin{equation}
\rho_{c}(t)=diag(|C_{0}^{\uparrow}|^{2},|C_{0}^{\downarrow}|^{2})\label{eq:SLlarge}
\end{equation}
from which the minimum value of the entropy can finally be obtained:
\begin{equation}
	S_\mathrm{min}=-|C_{0}^{\uparrow}|^{2} \log_2 |C_{0}^{\uparrow}|^{2} - |C_{0}^{\downarrow}|^{2} \log_2 |C_{0}^{\downarrow}|^{2}~.
\end{equation}

\end{document}